\documentclass[
    prl,                
    twocolumn,          
    superscriptaddress, 
    floatfix,           
    showkeys,           
    amsmath,amssymb,    
    longbibliography    
]{revtex4-2}

\usepackage{graphicx}    
\usepackage{dcolumn}     
\usepackage{bm}          
\usepackage[version=4]{mhchem} 
\usepackage{siunitx}     
\usepackage[colorinlistoftodos]{todonotes}

\usepackage{color}       
\usepackage{hyperref}    

\begin{document}

\title{Unveiling Micrometer-Range Spin-Wave Transport in Artificial Spin Ice}

\author{Syamlal Sankaran Kunnath}
\email{syamlal.sankaran@amu.edu.pl}
\affiliation{Institute of Spintronics and Quantum Information, Faculty of Physics and Astronomy, Adam Mickiewicz University, 61-614 Poznan, Poland}

\author{Mateusz Zelent}
\affiliation{Institute of Spintronics and Quantum Information, Faculty of Physics and Astronomy, Adam Mickiewicz University, 61-614 Poznan, Poland}
\affiliation{Fachbereich Physik and Landesforschungszentrum OPTIMAS, Rheinland-Pfälzische Technische Universität
Kaiserslautern-Landau, 67663 Kaiserslautern, Germany}

\author{Pawel Gruszecki}
\affiliation{Institute of Spintronics and Quantum Information, Faculty of Physics and Astronomy, Adam Mickiewicz University, 61-614 Poznan, Poland}

\author{Maciej Krawczyk}
\affiliation{Institute of Spintronics and Quantum Information, Faculty of Physics and Astronomy, Adam Mickiewicz University, 61-614 Poznan, Poland}

\date{\today}

\begin{abstract}

Artificial spin ice (ASI) systems exhibit fascinating phenomena, such as frustration and the formation of magnetic monopole states, and Dirac strings. However, exploring the wave phenomena in these systems is elusive due to the weak dipolar coupling that governs their interactions. In this study, we demonstrate coherent spin-wave propagation in an hybrid ASI system, which is based on a multilayered ferromagnetic thin film with perpendicular magnetic anisotropy and in-plane magnetized nanoelements embedded within it. We show that this system enables spin-wave transmission over a one-micrometer distance via exchange-mediated coupling between subsystems and evanescent spin-wave tunneling through the out-of-plane magnetized parts. This system overcomes the limitations of purely dipolar interactions in standard ASIs while preserving their fundamental properties. Thus, it provides a platform for studying spin-wave phenomena in frustrated ASI systems and paves the way for exploiting them in analog signal processing with spin waves.

\end{abstract}


\maketitle

\section{Introduction}

Magnonics---the study of spin waves (SWs) that carry angular momentum without net charge transport—is emerging as a key paradigm for wave-based computing in nanoscale~\cite{chumak2015magnon, barman20212021, chumak2022advances}. Patterned nanostrips~\cite{qin2021nanoscale, gruszecki2021local, lisiecki2019reprogrammability} and reconfigurable magnonic crystals~\cite{krawczyk2014review, haldar2016reconfigurable, wagner2016magnetic} serve as field-programmable SW platforms, suitable for analog computing. Artificial spin ice (ASI) systems, which comprise lithographically patterned arrays of monodomain nanomagnets arranged in a regular array, offer enormous reprogrammability~\cite{wang2006artificial, gilbert2014emergent, skjaervo2020advances}. It is, because when the nanomagnets interact via dipolar fields and the system exhibits magnetization microstate degeneracy at the vertices. This leads to a frustrated ground state and the emergence of magnetic charge physics at the vertices. 
Extensive characterization has revealed rich localized SW modes: edge, bulk and vertex~\cite{lendinez2019magnetization, gliga2020dynamics, jungfleisch2016dynamic, gliga2013spectral}, positioning ASI as a potential reconfigurable magnonic platform~\cite{gartside2018realization, gartside2022reconfigurable}.

Yet one critical capability of ASIs for magnonic applications remains unexplored. It is a \textit{long-range spin-wave transport across an ASI lattice}. Despite extensive characterization of localized resonances ~\cite{lendinez2019magnetization, gliga2020dynamics, jungfleisch2016dynamic, gliga2013spectral} and magnon-magnon coupling~\cite{montoncello2023brillouin, dion2024ultrastrong, sultana2025magnon}, 
to the best of our knowledge, no prior work has demonstrated the propagation of SWs across an ASI lattice. This is because standard ASI (s-ASI) systems considered so far, can only support limited SW transmission due to weak dipolar interactions, which hinder efficient energy transfer.

Recent work has demonstrated that embedding 2D in-plane magnetized ASI nanoelements in a multilayer with perpendicular magnetic anisotropy (PMA) creates strong exchange-mediated coupling between the SW modes of the ASI system and the PMA matrix~\cite{sankaran2024enhancement}. Building on this foundation, we now demonstrate via micromagnetic simulation that such a hybrid ASI--PMA architectures support long-distance SW transport spanning multiple lattice periods for two types of ASI modes: edge localized mode (EM) and bulk mode (BM).  Their decay lengths cross 1~µm and their group velocities reach hundreds of meters per second, under a modest out-of-plane bias magnetic field. 
The mechanism fundamentally differs from dipolar transfer in a s-ASI systems: ASI elements embedded in and exchange-coupled to the PMA film create an effective exchange-coupled background bridging gaps and supporting an evanescent SW channeling. Attenuation reflects controlled evanescent coupling tunable via vertex gaps or bias magnetic field. 

The proposed architecture is compatible with current nanofabrication techniques--focused-ion-beam (FIB) milling or direct-write annealing~\cite{beaujour2011ferromagnetic, pan2020edge, brock2024grayscale, das2024tuning}--enabling tailoring of PMA in thin ferromagnetic mulitlayers with high precision, down to tens of nanometers. Furthermore, since ASI microstates are reprogrammable by applied fields, these results establish ASI-PMA as a tunable magnonic platform where transmission length and active frequency band can be selected by the written magnetization microstates. Combined with ASI's reconfigurable SW mode spectrum, this suggests a path toward field-programmable magnonic circuits for neuromorphic computing~\cite{chumak2015magnon, barman20212021, chumak2022advances}. Moreover, the proposed ASI-PMA system paves the way for the study of the dynamics of emergent magnetic monopole states and their interactions via SW channeling along Dirac strings.

\section*{Results and discussions}

\begin{figure*}[t]
    \centering
    \includegraphics[width=1.0\linewidth]{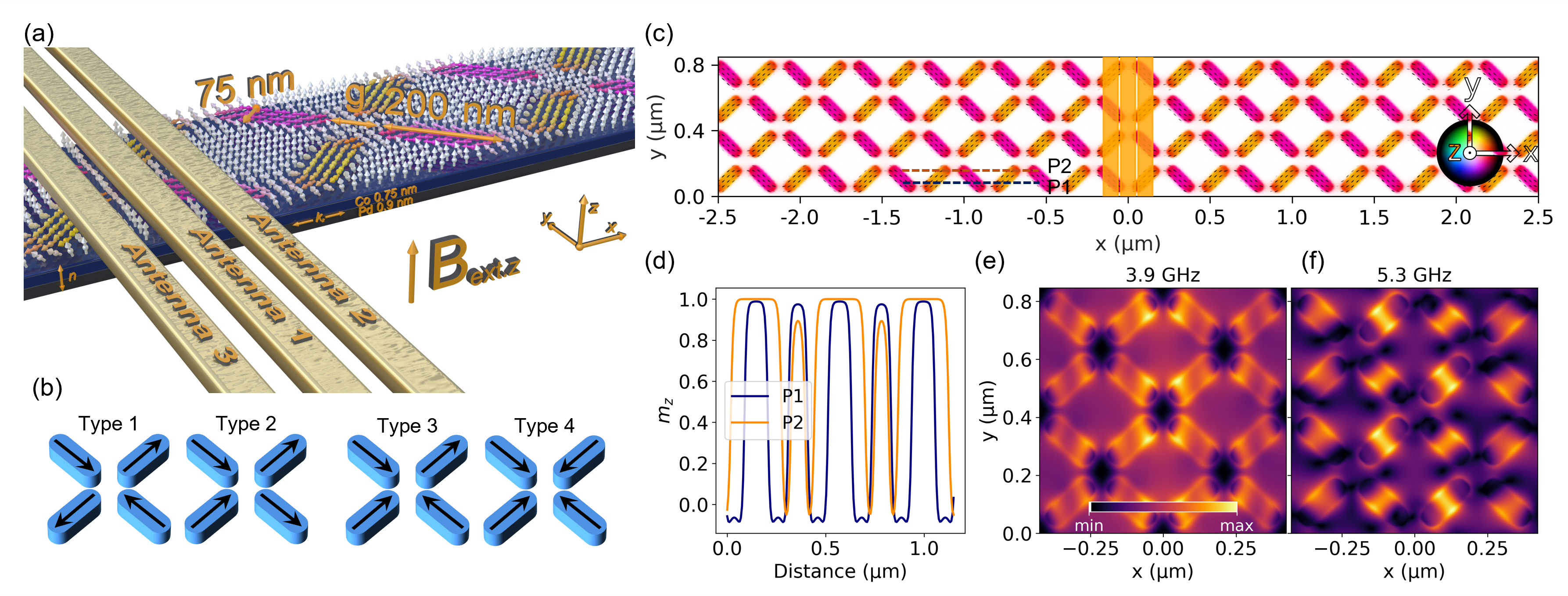}
    \caption{\textbf{ASI-PMA structure and resonant modes.} (a) Schematic of a 1D chain of square ASI-PMA lattice (nanoelements: 200 × 75 nm²) embedded in Co/Pd multilayer. Yellow regions indicate microwave antennas. (b) Four vertex types (Type 1–4) classified by spin configuration and net charge $Q$. (c) Relaxed magnetization in ASI-PMA ($-2.5$ to $2.5$ µm).The hue represents the in-plane orientation of the magnetization, while the brightness indicates the out-of-plane value (black: $-z$ and white: $+z$). (d) Out-of-plane magnetization ($m_z$) static magnetization cross-sections across the nanoelement core (P1) and edge (P2) positions. The orange vertical stripes mark the antenna used for the SW excitation. (e,f) Mode profiles (dynamic $m_z$ component) of the edge mode (EM, 3.9 GHz) and the bulk mode (BM, 5.3 GHz) spanning distance from $-0.42$ to 0.42 µm around the antenna.}
    \label{fig1}
\end{figure*}

The system under consideration is shown in Fig.~\ref{fig1}(a). It consists of a 12.73-µm long, 848.5 nm width waveguide of perpendicularly magnetized Co/Pd multilayer, in which an array of in-plane-mangetized  disconnected nanoelements (200 nm long and 75 nm wide)
are embedded to form a square ASI pattern (30 unit cells along the $x$-axis and two unit-cells along the $y$-axis with periodic boundary conditions along the $y$-axis), i.e., ASI-PMA system. The size and shape of the ASI nanoelements are chosen to maintain the monodomain state in the presence of moderate external magnetic fields and at remanence. The nanoelements are separated by 100 nm vertex gaps (edge-to-edge) and coupled through the PMA matrix via 90° domain walls at the ASI nanoelement-PMA interfaces. 
To model our system, we employ the effective medium approximation to model the Co/Pd multilayer as a homogeneous medium~\cite{Woo2016, Dhiman2024, Gieniusz2024}. We assume the saturation magnetization $M_{\text{s}} = 810$~kA/m, exchange constant $A_{\text{ex}} = 13$~pJ/m, anisotropy constant in PMA matrix $K_\text{u} = 4.5 \times 10^{5}$~J/m³, and damping $\alpha = 0.008$~\cite{sankaran2024enhancement, moalic2024role, pan2020edge}. The nanoelements have identical parameters, except for anisotropy, which is set to zero.
Micromagnetic simulations are performed using Amumax~\cite{amumax2023}, a modified MuMax3~\cite{vansteenkiste2014design} package (details in Supporting Information, Sec. I).

Before the dynamic simulations, the system is first relaxed to the ground state under the given applied bias magnetic field with an initial out-of-plane magnetization configuration. 
At remanence, PMA matrix relax to a domain structure, which causes unwanted SW scattering (see, Supporting Information Sec. III), therefore, an out-of-plane bias field ($B_{\text{ext},z}$) is applied to establish uniform PMA magnetization while preserving the ASI in-plane magnetized state.
Figure~\ref{fig1}(c) shows the relaxed magnetization configuration across the part of the simulated domain at $B_\text{ext}=50$~mT along the $z$-axis, and Fig.~\ref{fig1}(d) shows out-of-plane static magnetization component ($m_z$) maps of marked cross-sections. The lattice exhibits Type 2 of the ASI nanoelements vertex states (with total magnetic charge $Q=0$), a favorable configuration in ASI systems~\cite{skjaervo2020advances, gliga2020dynamics}. Importantly, there exists a strong dipolar interaction between nanoelements, as indicated by the magnetization tilt at the edges of the nanoelements, which is evident in the $m_z$ maps (see, Fig.~\ref{fig1}(c) and (d)). These interactions prove the existence of interactions in the ASI-PMA system similar to those in s-ASI. Also similarly to s-ASI systems, there are other possible magnetization configurations in ASI-PMA, as shown schematically in Fig.~\ref{fig1}(b). These are metastable states with higher energy (Type 3 with $Q=2$ and Type 4 with $Q=4$), and their stability has been confirmed by predefined magnetic states in relaxation simulations. In this paper, however, we only analyze the resonant SW modes and their propagation characteristics for Type 2.

\textbf{Spin-wave transmission in ASI lattices.} 

For the SW excitation we use a microwave magnetic field from three synchronized microwave stripline antennas (44.2 nm in width along the $x$-axis and  extended along the $y$-axis) positioned symmetrically across one unit cell, as shown schematically in Fig.~\ref{fig1}(a) and in Fig.~\ref{fig1}(c). Antenna 1 is at the center ($x = 0$ nm), Antennas 2 and 3 are at adjacent nanoelements ($x = \pm 106$ nm). The microwave magnetic field, in the form of a sinc function in the time domain, from all antennas oscillates in phase. This configuration minimizes scattering of excited waves, while ensuring full lattice coverage. To avoid SW reflections from the waveguide ends, we employ a 640 nm wide absorbing layers at the left and right sides of the waveguide where damping gradually increases up to 0.25 \cite{gruszecki2021local, gruszecki2022inelastic}.

The SW spectrum from the excitation area is rich with several intense peaks (see the Supporting Information, Sec. II, for the FFT spectra and SW profiles). However, only two SW modes dominate in propagation, as we will see below. It is the EM at 3.9 GHz and the BM at 5.3 GHz. These are visualized in Fig.~\ref{fig1}(e) and (f), respectively, with amplitude concentrated at the edges and bulk of the nanoelements. 
These modes appear well below the ferromagnetic resonance frequency of the PMA matrix (6.7 GHz, i.e., the frequency of the first mode with the amplitude concentrated in the matrix, see Supporting Information, Fig.~S1), nevertheless, they exhibit collective oscillations character. It is very similar to the observations in an s-ASI systems under ferromagnetic resonance conditions~\cite{gliga2013spectral, jungfleisch2016dynamic}. However, in our case, the SW intensity is significant not only under the antennae but also at large distances from them. In other words, SW propagation is preserved.

\begin{figure}[t]
    \centering
    \includegraphics[width=\linewidth]{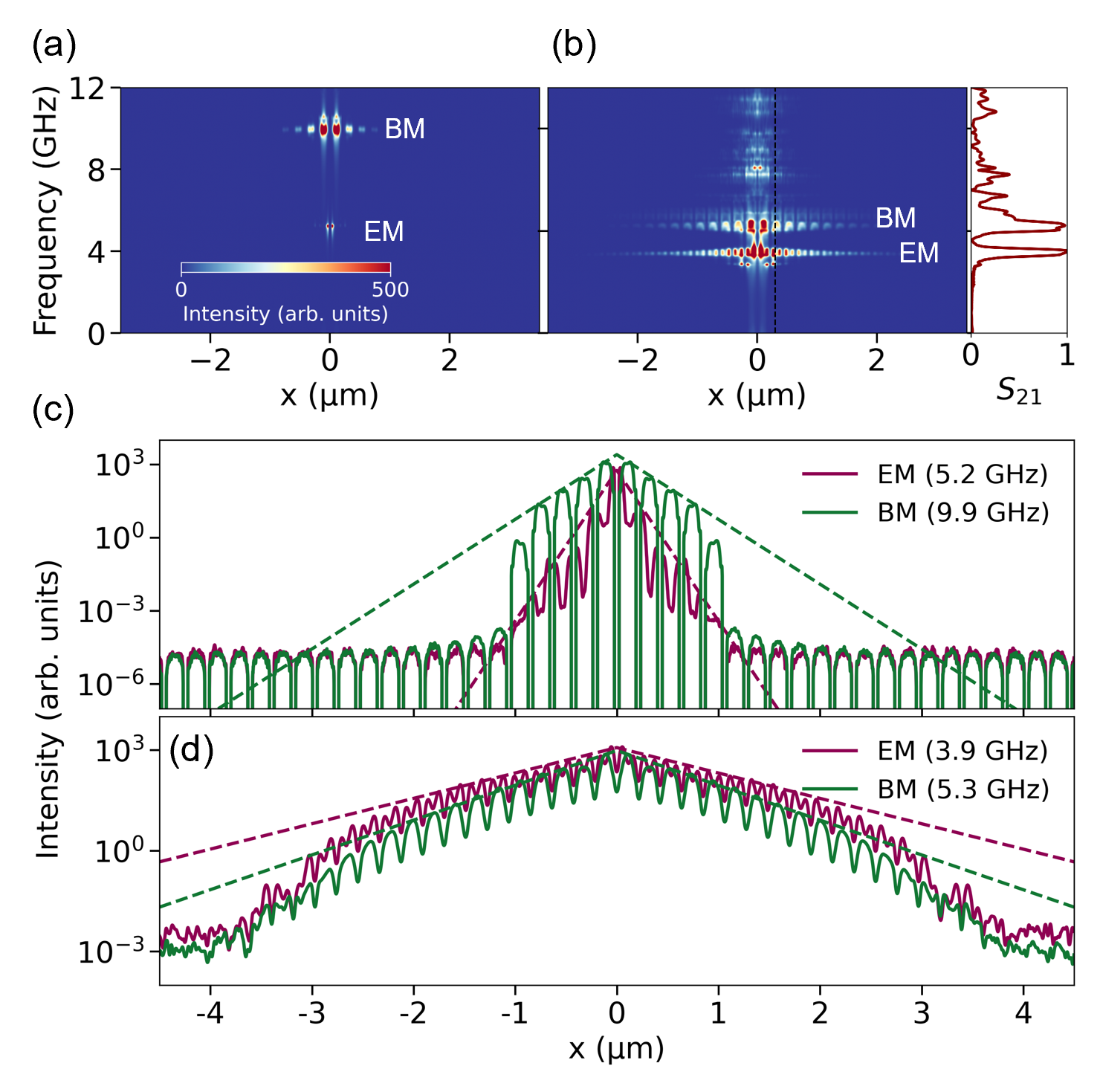}
    \caption{\textbf{Spin-wave transmission in s-ASI and ASI-PMA systems.} 2D spatial intensity maps (sum of the squared amplitude of $m_x$, $m_y$ and $m_z$ magnetization components) for (a) s-ASI and (b) ASI-PMA systems. Right: Normalized transmission spectra ($S_{21}$) measured at 0.3~µm from the antenna center. (c) and (d) Spatial decay of the intensity of the propagating SW modes along $\pm x$-directions on log scale, showing EM and BM for ASI-PMA (d), and analogous modes for s-ASI (c). Dotted lines: the exponential fits $\propto \exp(-2x/\xi)$. In s-ASI, no bias field, in ASI-PMA $B_{\text{ext},z} = 50$ mT, in both cases the identical excitation was used.
    } \label{fig2}
\end{figure}

The spatially dependent transmission spectra $A(x,f)$, obtained from the Fourier-transformed magnetization dynamics and integrated over the waveguide cross section (details in Supporting Information, Sec. I), are shown in Figs. 2(a) and 2(b). Clearly, SW propagation is effective only in ASI-PMA system. As already mentioned, only two of the SW modes from many excited under the antenna, i.e., EM and BM, are effectively transmitted for a distance exceeding a few hundred nm.
This is also revealed in the transmission spectra ($S_{21}$) measured at 0.3~µm from the antenna center (right panel in Fig.~\ref{fig2}(b)). Still there are multiple modes, however, most of them experienced a significant decrease in intensity. In contrast, the  EM and BM modes display strong signals on both antenna sides, indicating symmetric excitation and propagation along $\pm x$ (see Supporting Information, Sec. I for details).

To quantify the difference in propagation length between s-ASI and ASI-PMA systems, we plot $A(x,f)$ as a function of $x$ for selected modes in Figs.~\ref{fig2}(c) and (d), respectively. We use a logarithmic scale and from  the fitting the $A(x,f)$ maxima to $\propto \exp(-2x/\xi)$, we extract the propagation length, $\xi$,  i.e., the distance over which the SW intensity decreases by a factor of $e$. The fits yield $\xi = 1.2$ \textmu m  for the EM (3.9 GHz) and $0.9$ µm  for the BM (5.3 GHz) in ASI-PMA, and there is a negligible propagation in s-ASI ($\xi < 0.25$ µm). This 4--5-fold enhancement of the propagation length demonstrates that embedding ASI in a PMA matrix dramatically improves coherent SW transport. While shorter than that of a uniform Co/Pd film ($\xi = 1.4$ µm, see the analysis in Supporting Information, Sec. IV), the ASI-PMA hybrid uniquely combines geometry-tuned modes and reconfigurability  with efficient propagation, which is unattainable simultaneously in either a homogeneous PMA multilayer or s-ASI~\cite{haldar2016reconfigurable}. Although the damping assumed in our ASI–PMA system is slightly lower than the typical Co/Pd value, it is consistent with optimised systems and sufficient to capture the essential propagation physics. It is also an experimentally achieved value in optimised metallic PMA systems of other composition, such as FePd~\cite{zhang2020low}. Furthermore, recent experimental studies show that PMA in thin films of Bi-doped yttrium iron garnet can also be tuned with FIB~\cite{das2024tuning}, offering even lower damping, and thus increased $\xi$. Additional damping analysis is provided in Supporting Information, Sec. VI.

\begin{figure}
    \centering
    \includegraphics[width=0.8\linewidth]{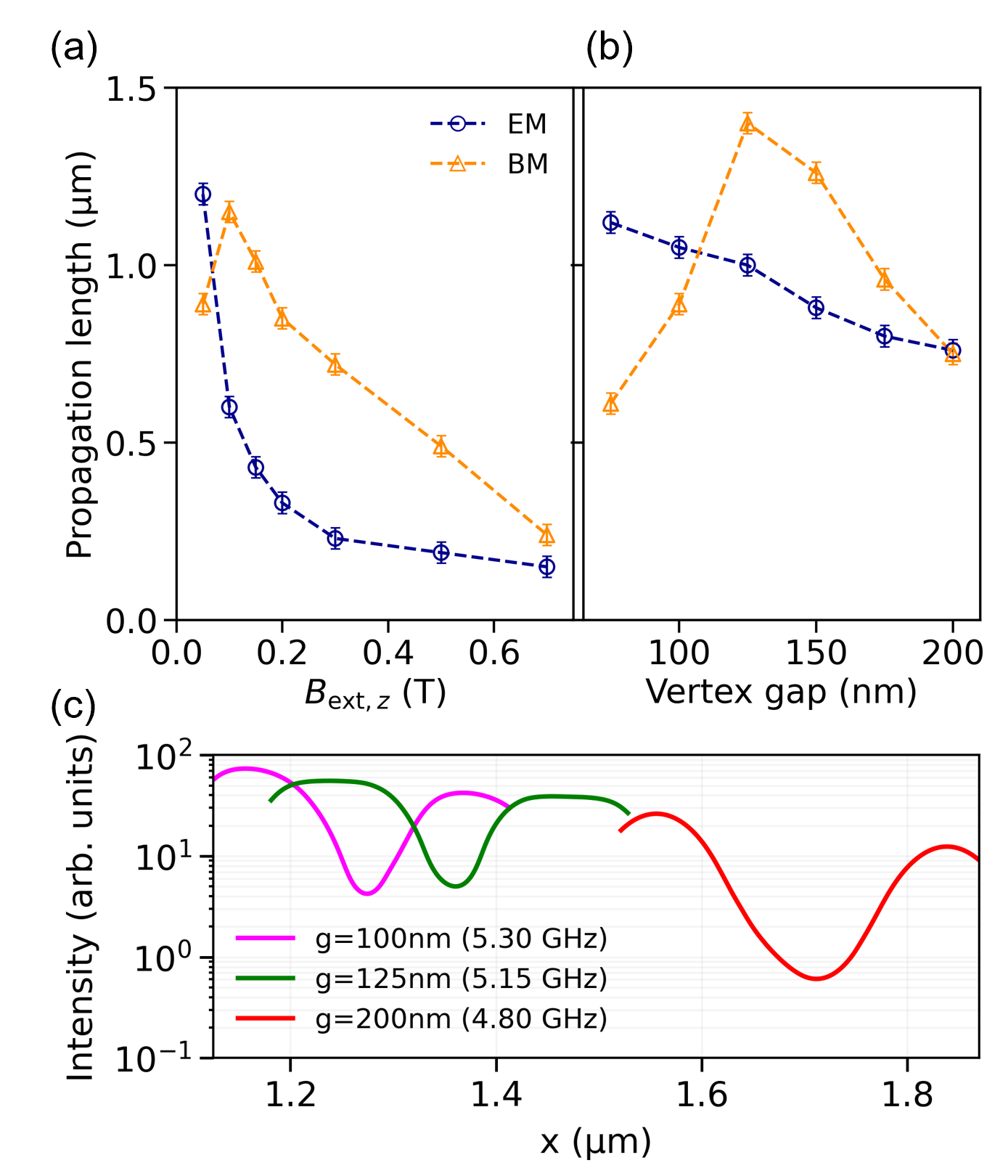}
    \caption{\textbf{Tunability and characteristics of propagating spin-waves in ASI-PMA.}  (a,b) Dependence of SW propagation length on bias field $B_{\text{ext},z}$ and vertex gap $g$, showing monotonic decay for EM and  maximum for BM. (c) Spatial distribution of the BM intensity, $A(x,f)$, through a single vertex gap (PMA matrix) and neighboring nanoelements for representative values of $g$.
    } \label{fig3}
\end{figure}

\textbf{Tunability, evanescence coupling, and dispersion relation}

Additional simulations allow to understand the significant enhanced propagation of SWs in ASI-PMA with respect to s-ASI. We conducted a systematic study of SW dynamics depending on the bias magnetic field and vertex gap size, $g$.
We found, see Fig.~\ref{fig3}(a) and (b), that the EM propagation length decreases \emph{monotonically} with both $B_{\text{ext},z}$ and $g$, whereas, unexpectedly, BM exhibits a \emph{non-monotonic} dependencies with a clear maxima at $B_{\text{ext},z}=0.1$~T with $\mathbf{\xi=\SI{1.2}{\micro\meter}}$ and at $g=\SI{125}{\nano\meter}$ with $\mathbf{\xi=\SI{1.4}{\micro\meter}}$. For completeness, we also evaluated the connected ASI-PMA geometry $(g=\SI{0})$, where the vertex-localized mode propagation length is about $\mathbf{\xi=\SI{1.4}{\micro\meter}}$ and for the corresponding BM $\mathbf{\xi=\SI{0.3}{\micro\meter}}$ (see, Supporting Information, Sec. IX).

The observed dependence $\xi(g)$ for the EM (Fig.~\ref{fig3}(b)) is understandable, since increasing the separation between the nanoelements naturally reduces their coupling. However, unlike s-ASI, ASI-PMA has a nonzero SW intensity in between the nanoelements (compare spectra Fig.~\ref{fig2}(c) and (d) and a drop in SW intensity in the vertex-gap areas). Since, the EM frequency is well below the ferromagnetic resonance of the PMA matrix, the SW in the PMA matrix exhibits \emph{evanescent wave} character with imaginary wave vectors and exponential spatial decay across the PMA gap. It mirroring tunneling transport through air-gap~\cite{matsumoto2020observation} and at material interfaces~\cite{verba2020spin}.
The evanescent coupling of SW modes via vertex gap exists also for BM. In Fig.~\ref{fig3}(c), we plot width-averaged SW intensity of this mode in a single unit-cell of ASI-PMA (for the EM, see Supporting Information, Sec. VII). At $g=125$ nm, where $\xi$ is maximal, the SW mode covers the largest part of the nanoelements (see the flat parts) and decays the least compared to other cases. We quantify the decay of the SW intensity in a PMA vertex gap as 
\begin{equation}
\mathrm{Decay}(g, f) = 20\log_{10}\!\left(\frac{A_{\mathrm{NM1}}(g,f)}{A_{\mathrm{NM2}}(g,f)}\right)\ \mathrm{dB},
\end{equation}
where, NM1 and NM2 are the neighboring ASI nanoelements 1 and 2, respectively, located at $\mathbf{\SI{1.0}{\micro\meter}}$ from the antenna along the wave-propagation path.
It shows an exponential decay of: $\,5.2~\mathrm{dB}$ at $g=\SI{100}{\nano\meter}$, $\,3.2~\mathrm{dB}$ at $g=\SI{125}{\nano\meter}$, and $\,6.5~\mathrm{dB}$ at $g=\SI{200}{\nano\meter}$. This attenuation correlates directly with the variation of the propagation length: $\xi$ increases by approximately 55\% between $g=\SI{100}{\nano\meter}$ and $g=\SI{125}{\nano\meter}$, and then decreases by 46\% when the gap is increased to $g=\SI{200}{\nano\meter}$.

\begin{figure}
    \centering
    \includegraphics[width=\linewidth]{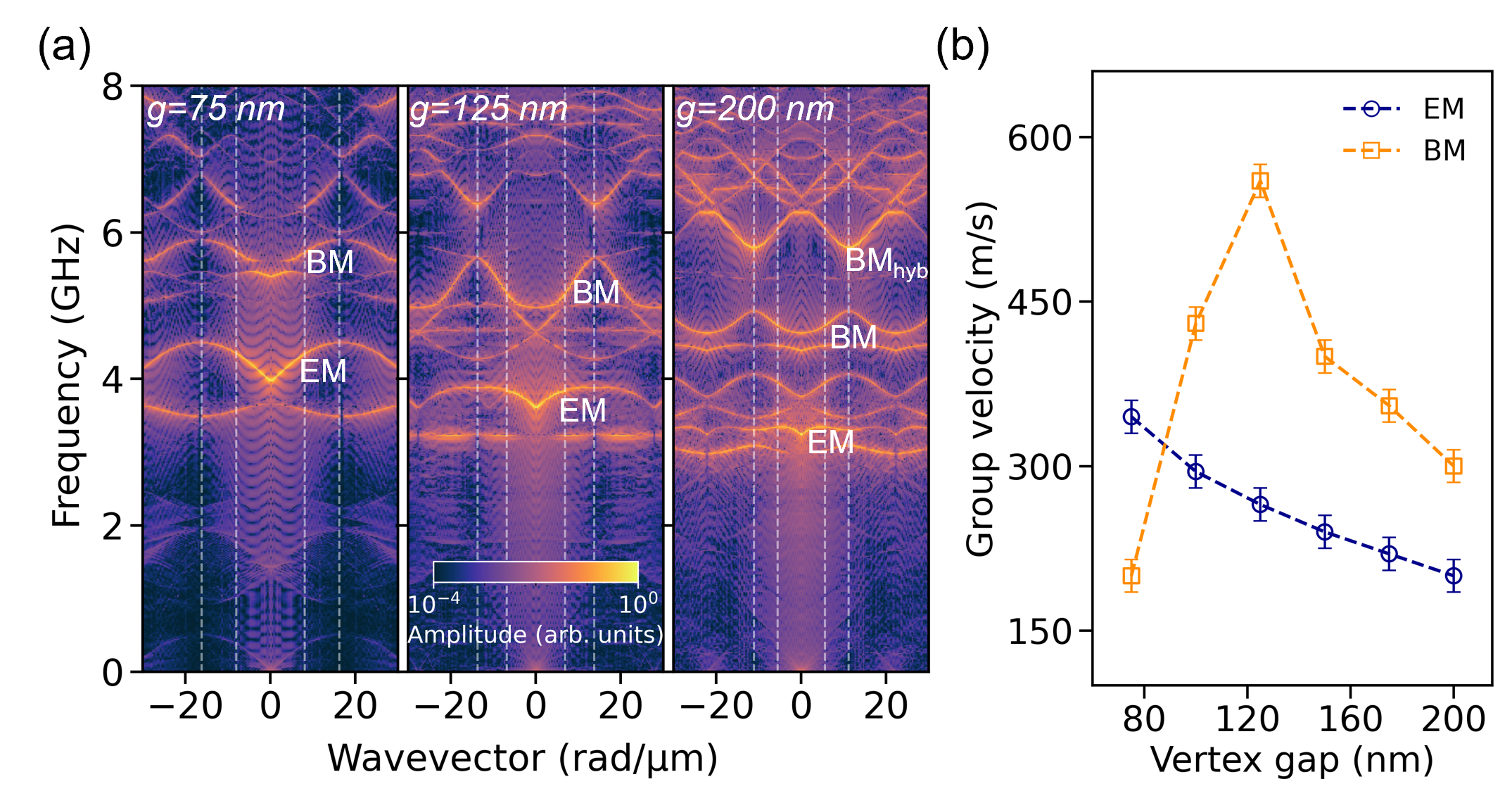}
    \caption{\textbf{Dispersion characteristics of propagating spin waves in ASI-PMA.} (a) Dispersion relations of SWs in ASI-PMA system at three selected values of $g$. Vertical dashed black lines mark the Brillouin-zone boundaries. (b) Maximum group velocity $v_\text{g}^{\max}$ derived from despersion relations for the EM and BM bands. All simulations made at 50~mT bias magnetic field.
    } \label{fig4}
\end{figure}

A change in $g$ also results in changes to other aspects of the system, such as the dispersion relation of SWs and the underlying change in the static magnetization configuration. Figure~\ref{fig4}(a) presents the dispersion relation for $g=\{75,125,200\}\,\mathrm{nm}$ (for comparison, the homogeneous PMA film dispersion is shown in Supporting Information, Sec. IV). There is a systematic decrease in the frequency of the EM and BM bands as $g$ increases. With this change, the EM bandwidth decreases, but the change of the BM bandwidth is not fully clear due to hybridization and crossing with other modes.
We extract a group velocity from the dispersion relation, as $v_\text{g} = 2\pi\,\Delta f/\Delta k$, where $\Delta k \equiv k_2-k_1$ and $\Delta f \equiv f(k_2)-f(k_1)$. 

Figure~\ref{fig4}(b) reports the \emph{maximum} group velocities $v_\text{g}^{\max}$ as a function of $g$ for EM and BM, computed by scanning each mode branch in the first Brillouin zone and selecting the $k$-interval $[k_1,k_2]$ that maximizes $\Delta f/\Delta k$, both taken \emph{along a single monotonic segment} of the targeted dispersion branch.

The narrowing monotonically of the EM bandwidth with $g$, yields $v_\text{g}^{\max} \approx \SI{350}{\meter\per\second}$ at $g=\SI{75}{\nano\meter}$, $\SI{330}{\meter\per\second}$ at $\SI{100}{\nano\meter}$, and lower for larger $g$, which is consistent with a monotonic decline of $\xi$ with increasing $g$ shown in Fig.~\ref{fig3}(b).
For BM, gap-driven reconfiguration of the band structure yields a \emph{maximum} of $v_\text{g}^{\max}=\mathbf{\SI{560}{\meter\per\second}}$ at $g=\SI{125}{\nano\meter}$. Despite the $\sim\SI{5.0}{dB}$ tunneling loss at this value of the vertex gap, the elevated $v_\text{g}^{\max}$ yields a maximum of the decay length, $\mathbf{\xi=\SI{1.4}{\micro\meter}}$, as shown in Fig.~\ref{fig3}(b).
For smaller or larger $g$ values, the bandwidth and $v_\text{g}^{\max}$ recede, and $\xi$ drops accordingly.

Turning to the changes in the static magnetization configuration caused by the change in in the vertex gap. As $g$ increases, the static dipolar interactions between neighboring nanoelements weaken, resulting in a transformation from an S-state magnetization configuration to saturation in individual nanoelements (see the detailed analysis in Supporting Information, Sec. VIII). In the PMA matrix, we also observe a systematic increase in magnetization alignment along the $z$-axis as $g$ increases, demonstrating the decreasing influence of neighboring in-plane magnetized nanoelements. 
However, since these changes are monotonic with increasing $g$, they do not explain the observed maximum in penetration length or group velocity for BM. Nevertheless, the changes in the static magnetic configuration can be responsible for the maximum in propagation length of BM with respect to the bias magnetic field, as observed in Fig.~\ref{fig3}(a). At this field value, the bias magnetic field, oriented along the $+z$ direction, compensates for the stray magnetic field from the PMA matrix oriented in the opposite direction within the nanoelement areas (see, Supporting Information, Sec. V for details). Thus, the static magnetization in the ASI nanoelements is oriented in the film plane in the central part at $B_{\text{ext},z} \approx 100$ mT providing optimal conditions for the evanescent coupling via the PMA matrix. 

We can conclude that the net propagation length results from superposing two effects: (i) the shared evanescent penalty, which worsens with increasing $g$, and (ii) the mode-specific change in the band structure with $g$.
For EM, both effects align (bandwidth and $v_\text{g}^{\max}$ down, decay up) $\Rightarrow$ \emph{monotonic} decline.
For BM, improved dispersion at $g=\SI{125}{\nano\meter}$, which aligns with the reduced evanescent loss at this spacing, sets the optimum at $B_{\text{ext},z}=\SI{50}{\milli\tesla}$, $g=\SI{125}{\nano\meter}$. 

These trends highlight substantial optimization potential via band-structure engineering in PMA--ASI hybrids: by co-tuning $g$ and $B_{\text{ext},z}$ to reshape dispersion while constraining evanescent loss, one can systematically extend propagation length and enable on-chip SW routing.

\section*{Conclusions}

We demonstrate via micromagnetic simulations the \emph{ viable design} of the planar ASI system for long-range SW transport. By embedding square lattice of in-plane magnetized ASI nanoelements in perpendicularly magnetized film, we overcome the fundamental limitation of weak dipolar coupling 
that has precluded practical SW propagation in s-ASI systems. The hybrid ASI-PMA architecture establishes exchange-mediated evanescent channels, enabling propagation lengths of two types of SW modes, EM and BM, approaching 1.4~µm, i.e., a 5--6-fold improvement over s-ASI ($< 0.25$~µm). 
Furthermore, our finding indicates that the BM exhibits a \emph{non-monotonic dependence} as a function of the vertex gap, with the maximum of $\xi = 1.4$~µm and $v_\text{g}^{\max}$ = 560~m/s at $g = 125$~nm, $B_{\text{ext},z}=50$~mT. It reveals a profound trade-off: while larger vertex gaps increase evanescent attenuation, optimized band-structure at intermediate gap scales can compensate it, yielding net SW intensity enhancement. This counterintuitive behavior demonstrates that ASI propagation is not just a scaled-down version of homogeneous PMA films. Rather, it is a complex wave phenomenon governed by the interplay between geometry and evanescence, dipolar coupling,  and exchange interactions.

The fabrication of proposed ASI-PMA hybrid systems is compatible with current nanofabrication tools since the structure can be based on standard metallic multilayers with PMA and an ASI structure that is either imprinted by FIB or created through direct-write annealing. Characterization of the SW properties also can be done with existing techniques, like a µ-BLS, TR-MOKE or all-electric SW spectroscopy\cite{haldar2016reconfigurable,Qin2022,liu2018long,Vanatka2021}. Moreover, further material optimization, it is using low-damping ferromagnetic dielectric films, e.g. YIG, TmIG, promises propagation lengths exceeding several ~\textmu m \cite{qin2021nanoscale, timalsina2024mapping}, bringing ASI-based magnonic circuits toward practical neuromorphic computing architectures.

Since, the ASI microstates are reprogrammable via bias magnetic field, the transmission band and propagation length can also be tuned that way. This enables the first \textit{tunable magnonic platform} combining geometry-defined mode localization with field-switchable propagation—a capability absent so far in magnonic crystals or homogeneous films. Moreover, the presence of reprogrammable lattices, such as magnetic monopoles and Dirac strings provides an additional degree of freedom to tailor and guide SW transmission on demand\cite{gliga2013spectral,bhat2016magnetization}.
These results establish the ASI-PMA as a viable, reconfigurable platform for studying unexplored wave phenomena in frustrated systems, as well as on-chip SW routing and signal processing—bridging decades of ASI magnetization studies with the emerging field of wave-based computation.

\section*{Acknowledgments}
The study has received financial support from the National Science Centre of Poland, Grant No. UMO-2020/37/B/ST3/03936. The simulations were partially performed at the Poznan Supercomputing and Networking Center (Grant No. PL0095-01). MZ acknowledges that this, this project has received funding from the European Union´s Framework Programme for Research and Innovation HORIZON-MSCA-2024-PF-01 under the Marie Sklodowska-Curie Grant Agreement Project 101208951 – CNMA. 

\section{Competing interests}
The authors declare no competing interests.

\bibliographystyle{apsrev4-2}
\bibliography{bibliography}

\end{document}


\title{Supporting information: Unveiling Micrometer-Range Spin-Wave Transport in Artificial Spin Ice}

\author{Syamlal Sankaran Kunnath}
\email{syamlal.sankaran@amu.edu.pl}
\affiliation{Institute of Spintronics and Quantum Information, Faculty of Physics and Astronomy, Adam Mickiewicz University, 61-614 Poznan, Poland}

\author{Mateusz Zelent}
\affiliation{Institute of Spintronics and Quantum Information, Faculty of Physics and Astronomy, Adam Mickiewicz University, 61-614 Poznan, Poland}
\affiliation{Fachbereich Physik and Landesforschungszentrum OPTIMAS, Rheinland-Pfälzische Technische Universität
Kaiserslautern-Landau, 67663 Kaiserslautern, Germany}

\author{Pawel Gruszecki}
\affiliation{Institute of Spintronics and Quantum Information, Faculty of Physics and Astronomy, Adam Mickiewicz University, 61-614 Poznan, Poland}

\author{Maciej Krawczyk}
\affiliation{Institute of Spintronics and Quantum Information, Faculty of Physics and Astronomy, Adam Mickiewicz University, 61-614 Poznan, Poland}

\maketitle
\section{Micromagnetic simulations}

The micromagnetic simulations are performed using modified version of $Mumax^{3}$ \cite{vansteenkiste2014design} -- $Amumax$~\cite{amumax2023, sankaran2024enhancement}, which solves numerically the Landau--Lifshitz--Gilbert equation:
\begin{equation}
 \frac{\text{d}\mathbf{m}}{\mathrm{d}t}= 
 \frac{\gamma \mu_0}{1+\alpha^{2}} \left(\mathbf{m} \times \mathbf{H}_{\mathrm{eff}} + 
 \alpha  \mathbf{m} \times 
 (\mathbf{m} \times \mathbf{H}_{\mathrm{eff}}) \right),
\end{equation}
where $\textbf{m} = \textbf{M} / M_{\mathrm{S}}$ is the normalized magnetization, $\textbf{\text{H}}_{\mathrm{eff}}$ is the effective magnetic field acting on the magnetization, $\gamma=187$ rad/(s$\cdot$T) is the gyromagnetic ratio, $\mu_0$ is the vacuum permeability and $\alpha$ is gilbert damping.
The following components are considered in $\textbf{H}_{\mathrm{eff}}$: demagnetizing field $\textbf{\text{H}}_{\mathrm{d}}$, exchange field $\textbf{\text{H}}_{\mathrm{exch}}$, uniaxial magnetic anisotropy field $\textbf{\text{H}}_{\mathrm{anis}}$, and external magnetic field $\textbf{\text{H}}_{\mathrm{ext}}$, and thermal effects have been neglected:
\begin{equation}
  \textbf{H}_{\mathrm{eff}} =
  \textbf{H}_{\mathrm{d}} + \textbf{H}_{\mathrm{exch}} + \textbf{H}_{\mathrm{ext}} + \textbf{H}_{\mathrm{anis}} +\textbf{h}_{\mathrm{mf}}.
\end{equation}
The last term, $\textbf{h}_{\mathrm{mf}}$ is a microwave magnetic field used for SW excitation. The exchange and anisotropy fields are defined as
\begin{equation}
  \textbf{H}_{\mathrm{exch}} = \frac{2A_{\mathrm{ex}}}{\mu_0 M_{\mathrm{S}}} \Delta \textbf{m},\;
  \textbf{H}_{\mathrm{anis}} =
\frac{2K_{\mathrm{u,bulk}}}{\mu_0 M_{\mathrm{S}}} m_z \hat{\textbf{z}},
\label{Eq:Fields}
\end{equation}
where $A_{\mathrm{ex}}$ is the exchange constant, $M_{\mathrm{S}}$ is the saturation magnetization, and $K_{\mathrm{u,bulk}}$ is bulk magnetic anisotropy constant.

The unit cell of the ASI-PMA system is discretized into a $5760 \times 384$ grid of cuboids in the $x$ and $y$ directions, with each unit cell measuring approximately $2.21 \times 2.21 \times$ [Co/Pd thickness]~nm$^3$. The chosen cell size is smaller than the exchange length (5.62~nm), ensuring numerical accuracy. Periodic boundary conditions with 8 repetitions are applied along both $x$ and $y$ directions. While the lattice constant is varied for the subsequent cases (Fig.~3 and 4 in the main text), the discretization remains fixed throughout the simulations. To avoid SW reflections from the waveguide ends, we employ 640-nm-wide absorbing boundary layers (ABLs) on both sides of the waveguide, where the Gilbert damping is gradually increased from its base value to 0.25 following a normalized quadratic profile $\alpha(i) = \alpha_{0} + 0.25\left(\frac{i}{\mathrm{grad}}\right)^{2}$, where $i$ indexes the discretization cell within the ABL and $grad$ denotes the total number of numerical steps used to span the ABL width. To avoid numerical artifacts originating from exact lattice symmetry, a small tilt of \(0.0001^\circ\) was introduced to the external field with respect to the \(z\)-axis, in all simulations.

The stabilization of the magnetization configuration is obtained in the following steps. For the ASI-PMA structure, initially, the magnetization of each cell in the PMA bulk part is aligned along the $z$-axis, while the ASI nanoelements are kept along the x-axis. After the relaxation, the ASI stabilizes into Type 2 vertex state. The distinct magnetization configurations such as monopole and anti-monopole vertex states were achieved by initializing the magnetization in a predetermined arrangement corresponding to the desired type, followed by a relaxation process. 

Subsequently, we excite the SWs with three microwave strip antenna with microwave magnetic field along the $x$-axis with a cut-off frequency of 12 GHz, and a peak amplitude of $1.25 \times 10^{-2}$~T. Such a type of antenna enables us to focus on the propagation of the SWs with the amplitude concentrated within the nanoelements as well as in an out-of-plane magnetized region. The magnetization dynamics were sampled at intervals of 27.77~ps over a period of 100~ns. 

\begin{figure*}
    \centering
    \includegraphics[width=1.0\linewidth]{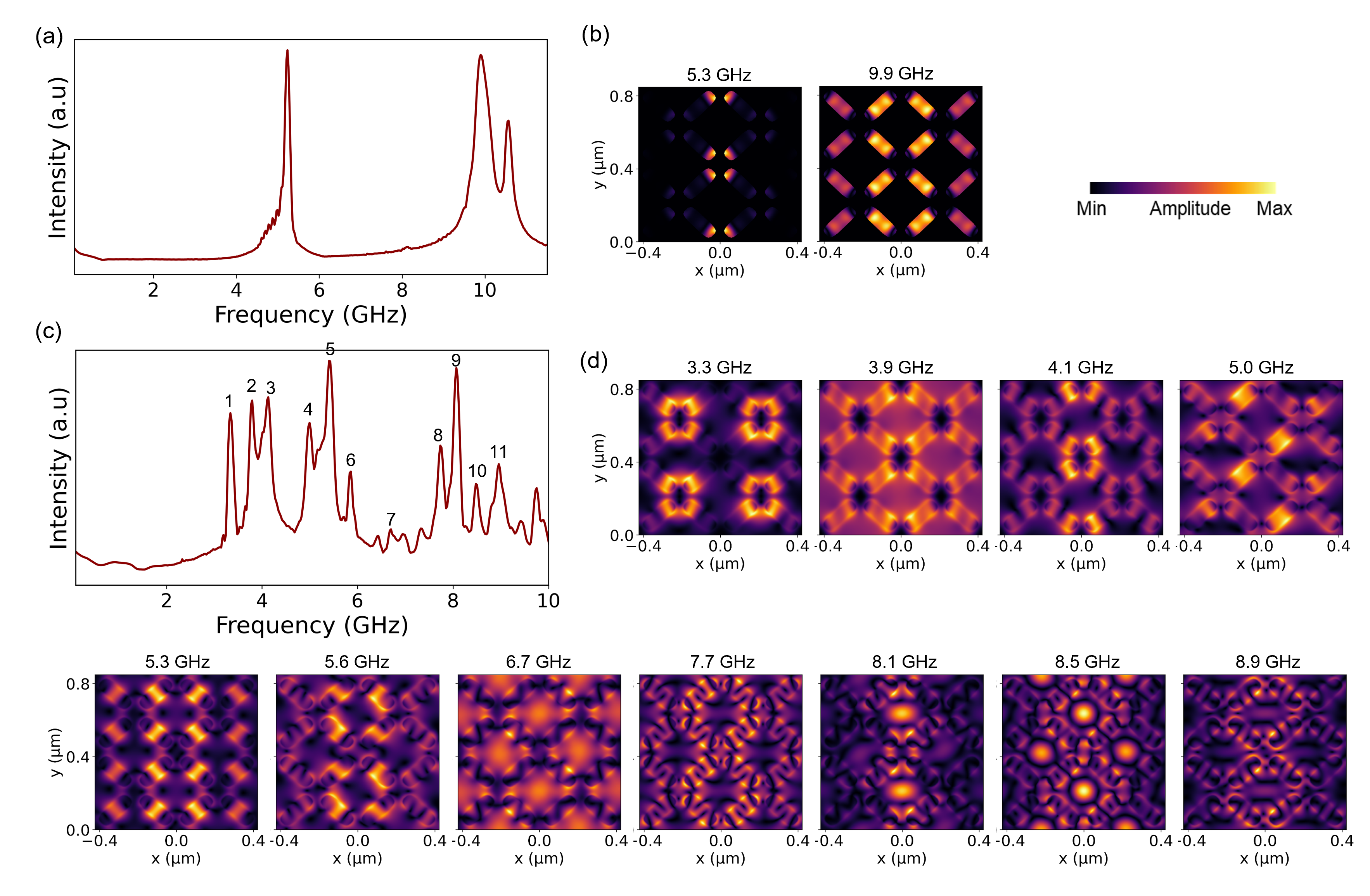}
    \caption{(a) The SW spectra ($m_x$ component) of s-ASI lattices in the absence of any applied field, (b) the corresponding amplitude of EM and BM. Similarly, (c) the spectra and (d) mode profiles of ASI-PMA lattices under $B_{\text{ext},z} = 50$ mT.}
    \label{figS1}
\end{figure*}

\begin{figure*}
    \centering
    \includegraphics[width=0.90\linewidth]{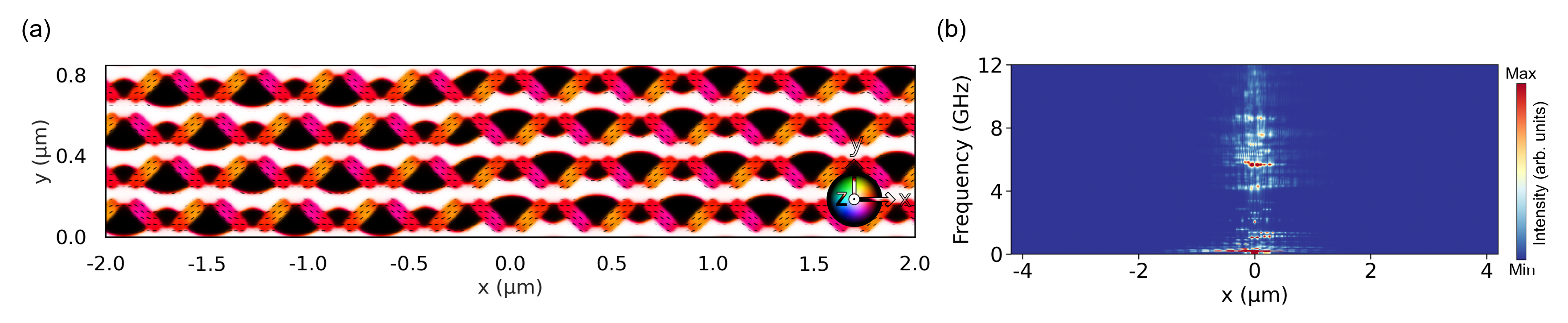}
    \caption{(a) Relaxed magnetization configuration of ASI-PMA lattices in the absence of bias magnetic field. The white and black regions corresponds to $+z$ and $-z$ orientation of the static magnetization, respectively. (b) 2D spatial map of the SW intensity for ASI-PMA lattices without any bias field.}
    \label{figS2}
\end{figure*}

\subsection*{Spin-wave transmission spectrum analysis}

To obtain the SW transmission spectrum, we used the space- and time-resolved magnetization from the micromagnetic simulations. A Hanning window was applied along the time axis to reduce spectral leakage, and a real-valued fast Fourier transform (FFT) was computed at each spatial cell. From the Fourier components, the local SW intensity was evaluated as
\[
I(x,y,f)=|m_x(x,y,f)|^{2}+|m_y(x,y,f)|^{2}+|m_z(x,y,f)|^{2}.
\]

To construct the two-dimensional transmission map (manuscript Fig. 2 (a, b)), the FFT of the time-dependent magnetization was performed at every spatial position along the waveguide. The position–frequency distribution was obtained from

\begin{equation}
A(x,f) = \sum_{y} I(x,y,f),
\end{equation}

The function \(A(x,f)\) represents the SW transmission spectrum, and positive (negative) \(x\) corresponds to propagation to the right (left) of the antenna. The resulting colormap \(A(x,f)\) provides the full frequency-dependent evolution of SW intensities along the structure. 

\subsection*{Dispersion relation spectrum analysis}

To obtain a clearer dispersion relation within the waveguide, simulations were performed on extended systems comprising 45 ASI lattice periods along the $x$-axis. The total lengths of the structures with vertex gaps of 100, 125, and 200 nm were 19.09, 20.68, and 25.45 µm, respectively, while maintaining same discretization cell size across all cases. Since the damping parameter does not influence the dispersion characteristics, a negligigble value of damping ($\alpha = 1\times10^{-7}$) was used. The magnetization dynamics were recorded up to a cut-off frequency of 8 GHz, with temporal sampling at 41.7 ps intervals over a total simulation time of 62.6 ns.

\section{Spectra and mode profiles of spin waves in s-ASI and ASI-PMA systems}

Figure~\ref{figS1} presents the SW spectra of the s-ASI systems and ASI-PMA systems. In the s-ASI lattices, two primary SW modes can be distinguished [Fig.~\ref{figS1}(a, b)]: the edge mode ($\text{EM}$ - 5.3 GHz) and the bulk mode (BM - 9.9 GHz), with their amplitudes localized at the edges and the central area of the nanoelements, respectively. Notably,  for the ASI lattices embedded in PMA multilayers [Fig.~\ref{figS1}(c,d)], the spectra, under an external bias field of $B_{\text{ext},z} = 50$ mT, is a richer due to mode splitting and interaction of the nanoelements with the PMA matrix. The modes of the nanoelements shift towards lower frequencies, with strongly excited EM and BM modes emerging at 3.9 GHz and 5.3 GHz, respectively. The corresponding SW propagation dynamics for both systems are presented in Fig. 2 of the main manuscript.

\section{Spin-wave transmission in ASI-PMA lattices without any bias field}

In the absence of bias magnetic field, the magnetization in the PMA multilayers in the ASI-PMA systems breaks into domains within the PMA matrix (Fig.~\ref{figS2}(a)). Applying a microwave drive via the antenna to this domain configuration excites numerous modes that are spatially localized across the structure and exhibit negligible propagation, as evident in the 2D spatial map shown in Fig.~\ref{figS2}(b).

\section{Spin-wave transmission in the PMA multilayer}

\begin{figure}
    \centering
    \includegraphics[width=1.0\linewidth]{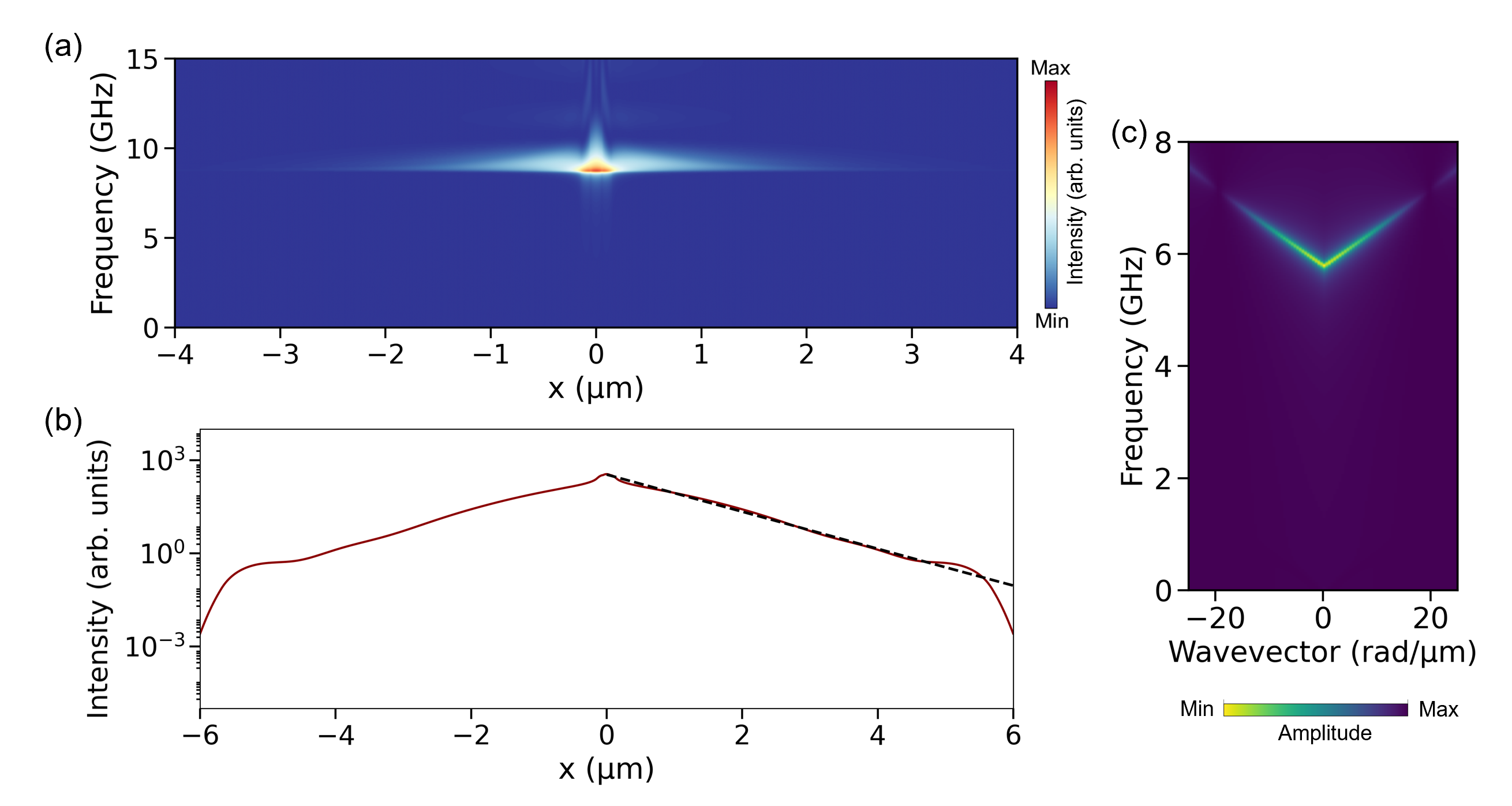}
    \caption{ (a) 2D spatial map of the SW intensity and (b) dependence of SW intensity (log-linear scale) corresponding to the propagation modes for a homogeneous PMA multilayers. The dotted line in (b) represents the fitting of simulation data by the exponential function. (c) The dispersion relation of SWs in PMA multilayer. In all simulations $B_{\text{ext},z}$ = 200 mT.}
    \label{figS3}
\end{figure}

Figure~\ref{figS3} presents the analysis of the SW propagation dynamics in homogeneous PMA multilayers (i.e., without ASI pattern), using dimensions consistent with the waveguide and nanomagnets shown in Fig.~2 of the main manuscript for direct comparison. As shown in Fig.~\ref{figS3}(a,b), the resonance mode of the PMA multilayer at 9.0 GHz exhibits propagation over a distance of 1.45 \textmu m. Note that this simulation was performed under an applied out-of-plane bias field of $B_{\text{ext},z}$ = 200 mT to avoid multi-domain formation and ensure a uniform magnetization state. The maximum group velocity ($v_\text{g} = 2\pi\,\Delta f/\Delta k$) of this resonance mode extracted from the dispersion relation shown in Fig.~\ref{figS3}(c) is 430 m/s.

\section{Magnetization maps under two different bias field values}

\begin{figure*}
    \centering
    \includegraphics[width=0.6\linewidth]{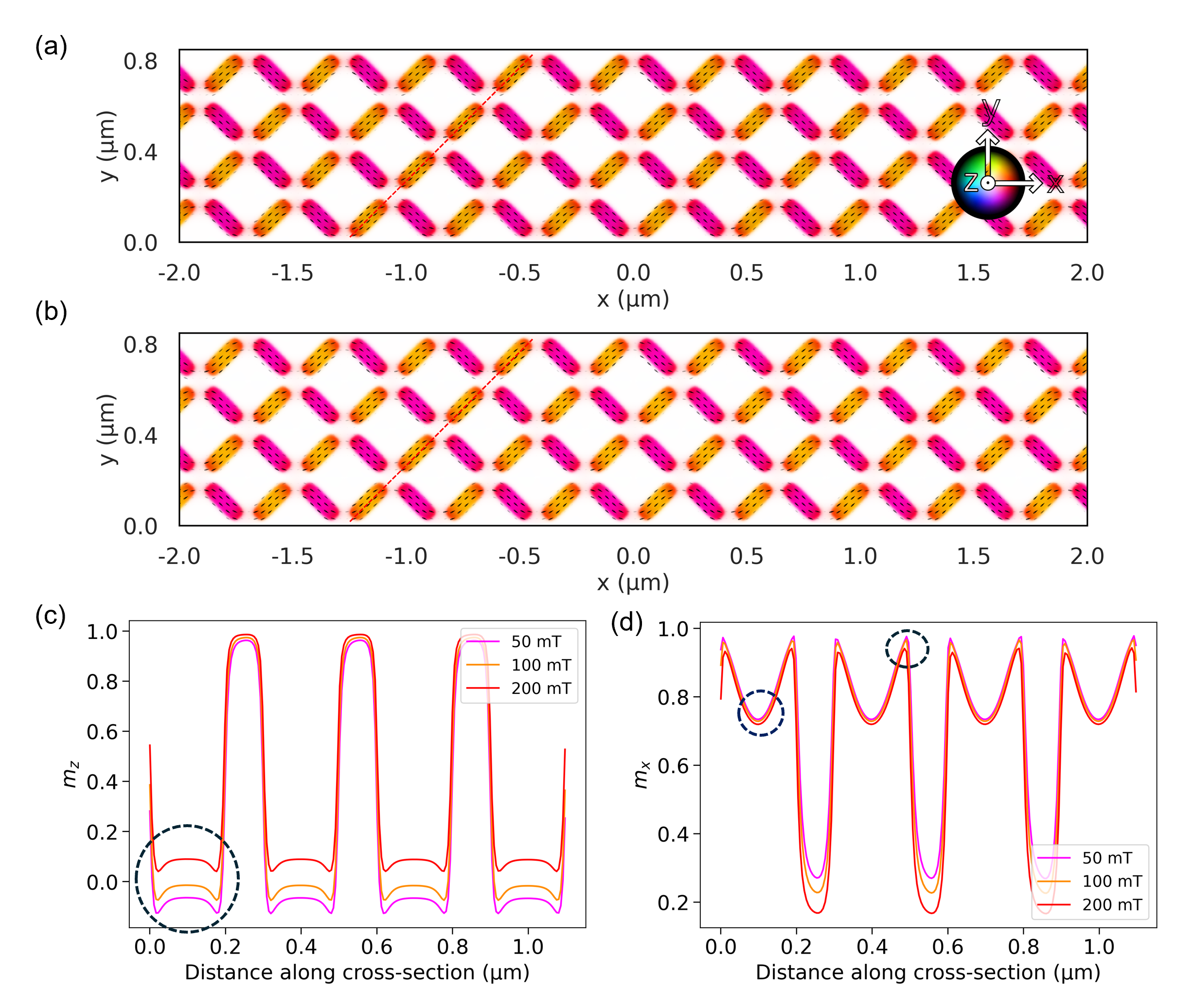}
    \caption{The static magnetization configuration of the ASI-PMA lattices under external out-of-plane fields of (a) $B_{\text{ext},z} = 50$ mT and (b) $100$ mT. (c) and (d) The cross-sectional profiles of the magnetization components, $m_x$ and  $m_z$, respectively, across the nanoelements for $B_{\text{ext},z} = 50$ mT, $100$ mT and $200$ mT. The plots are made along the red dashed lines indicated in panels (a) and (b).}
    \label{figS4}
\end{figure*}

\begin{figure}
    \centering
    \includegraphics[width=1.0\linewidth]{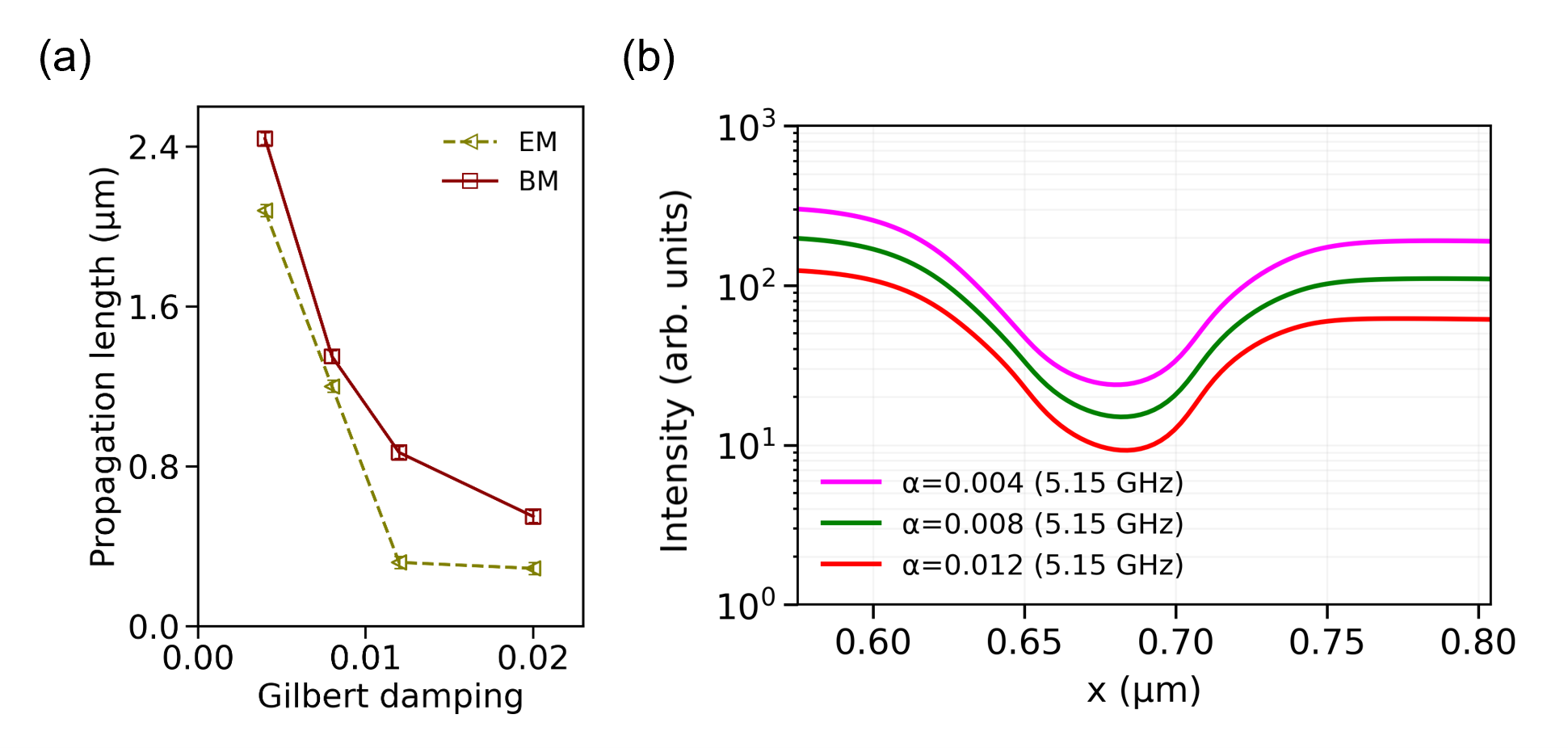}
    \caption{(a) Dependence of the SW propagation length on Gilbert damping constant in ASI-PMA lattice at magnetic field, $B_{\text{ext},z}$ = 50 mT and vertex gap $g$ = 125 nm. (b) Spatial decay of the intensity of the BM, measured along the propagation path ($+x$) from the center of one nanomagnet (NM1) to the next (NM2) through the PMA region at the vertex gap for the damping values, $\alpha$ = 0.004, 0.008 and 0.012.}
    \label{figS5}
\end{figure}

Figure ~\ref{figS4} (c, d) presents a comparative analysis of the cross-section of the static magnetization component maps shown in Figure ~\ref{figS4} (a, b), $m_x$ and $m_z$, for external out-of-plane magnetic fields of $B_{\text{ext},z} = 50$ mT, $100$ mT and $200$ mT across the nanomagnets. In all the cases, magnetic moment reorientation is clearly observed at the interfaces between the nanoelements and the PMA regions, transitioning from their intrinsic in-plane and out-of-plane alignments. Notably, the magnetization distribution exhibits a distinct lateral shift (highlighted by the black dotted line) in the $B_{\text{ext},z} = 100$ mT and $200$ mT case compared to $B_{\text{ext},z} = 50$ mT. This shift originates from the partial compensation of magnetization along the applied field direction within both the nanoelements and the PMA regions. 

\section{Effect of Gilbert damping constant on spin-wave transmission}

Figure~\ref{figS5}(a) illustrates the influence of the damping constant ($\alpha$) on the SW propagation length (all other material parameters are kept fixed). As expected, for the lowest damping value typically reported in FePd metallic multilayers ($\alpha$ = 0.004), the propagation length reaches $\mathbf{\xi=\SI{2.08}{\micro\meter}}$ for the EM mode and $\mathbf{\xi=\SI{2.44}{\micro\meter}}$ for the BM. Increasing the damping constant leads to a pronounced reduction in the propagation length, consistent with enhanced energy dissipation during wave propagation.
Figure~\ref{figS5}(b) presents the the intensity of SWs along the propagation direction of the BM, measured from the center of one nanomagnet (NM1) to the next (NM2) through the PMA matrix at a fixed vertex gap ($g$ = 125 nm) for three different damping values ($\alpha$ = 0.004, 0.008, 0.012).
Notably, the loss increases from 3.8 dB at $\alpha$ = 0.004 to 5.0 dB at $\alpha$ = 0.008, and further to 5.9 dB at $\alpha$ = 0.012.

\section{Evanescence wave coupling of EM through PMA matrix}

\begin{figure}
    \centering
    \includegraphics[width=0.75\linewidth]{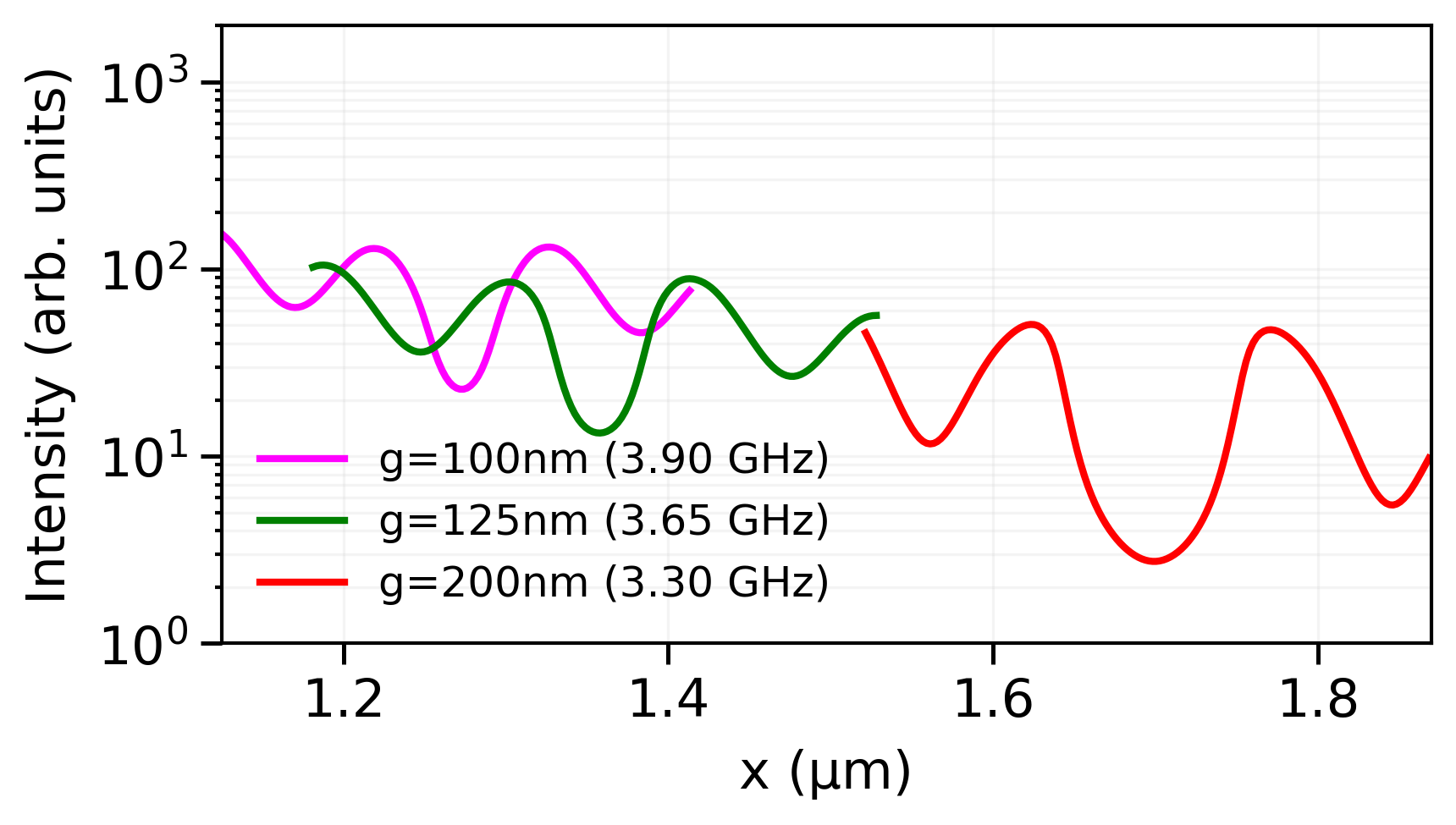}
    \caption{ Spatial distribution of the SW intensity of theEM, measured along the propagation path ($+x$) from the center of one nanomagnet (NM1) to the next (NM2) through the PMA region, for the vertex gaps, $g$ = 100 nm, 125 nm and 200 nm.}
    \label{figS6}
\end{figure}

Figure~\ref{figS6} presents the spatial distribution of the SW intensity along the propagation direction of theEM, measured from the center of one nanomagnet (NM1) to the next (NM2) through the PMA matrix for three different vertex gaps ($g$ = 100, 125 and 200 nm). The dependence is consistent with the trend observed for BM in Fig. 3 of the main text, the loss of EM increases from 0.7 dB at $g$ = 100 nm to 1.8 dB at $g$ = 125 nm, and further to 4.9 dB at $g$ = 200 nm.

\begin{figure*}
    \centering
    \includegraphics[width=0.80\linewidth]{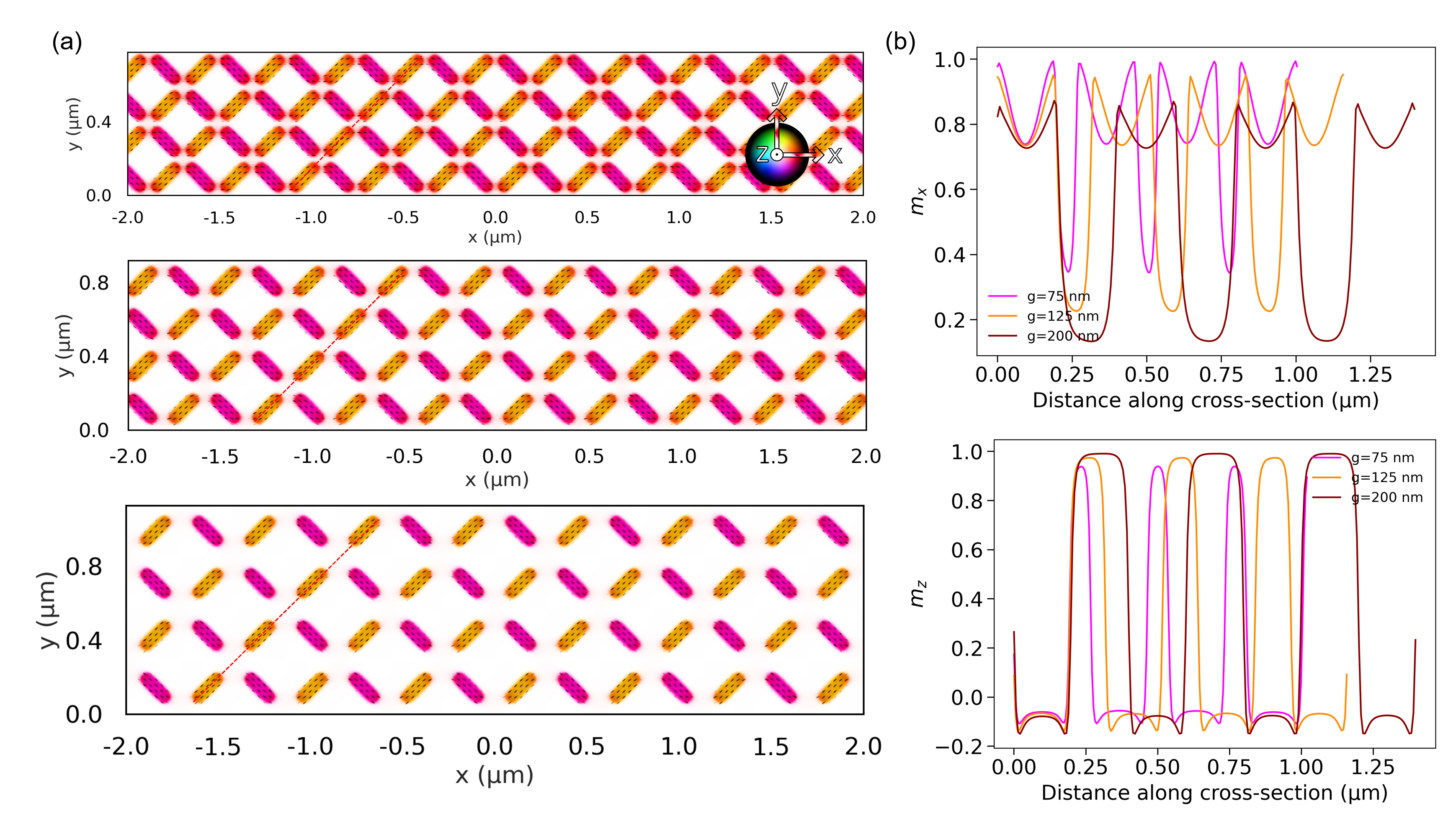}
    \caption{(a) The static magnetization configuration of ASI lattices under external out-of-plane fields of $B_{\text{ext},z} = 50$ mT, for vertex-gap width $g$ = 75 nm, 125 nm and 200 nm.  (b) Spatial variation of the in-plane magnetization component ($m_x$) along a cross-section marked by red-dashed lines shown in (a), for the different vertex-gap widths.}
    \label{figS7}
\end{figure*}

\section{Static magnetization variation with increasing the vertex-gap width}

Figure~\ref{figS7}(b) presents the spatial variation of the in-plane magnetization component ($m_x$) along the cross-section indicated in Fig.~\ref{figS7}(a) for vertex-gap sizes of $g =$ 75, 125, and 200 nm. As the gap increases, the static dipolar interactions between neighboring nanoelements weaken, resulting in a transformation from an S-state magnetization configuration to saturation in individual nanoelement. For the largest separation ($g = 200$ nm), the magnetization becomes nearly uniform, indicating minimal inter-element magnetic interaction.

\section{Spin-wave transmission through connected ASI-PMA lattices}

\begin{figure*}[t!]
    \centering
    \includegraphics[width=1.0\linewidth]{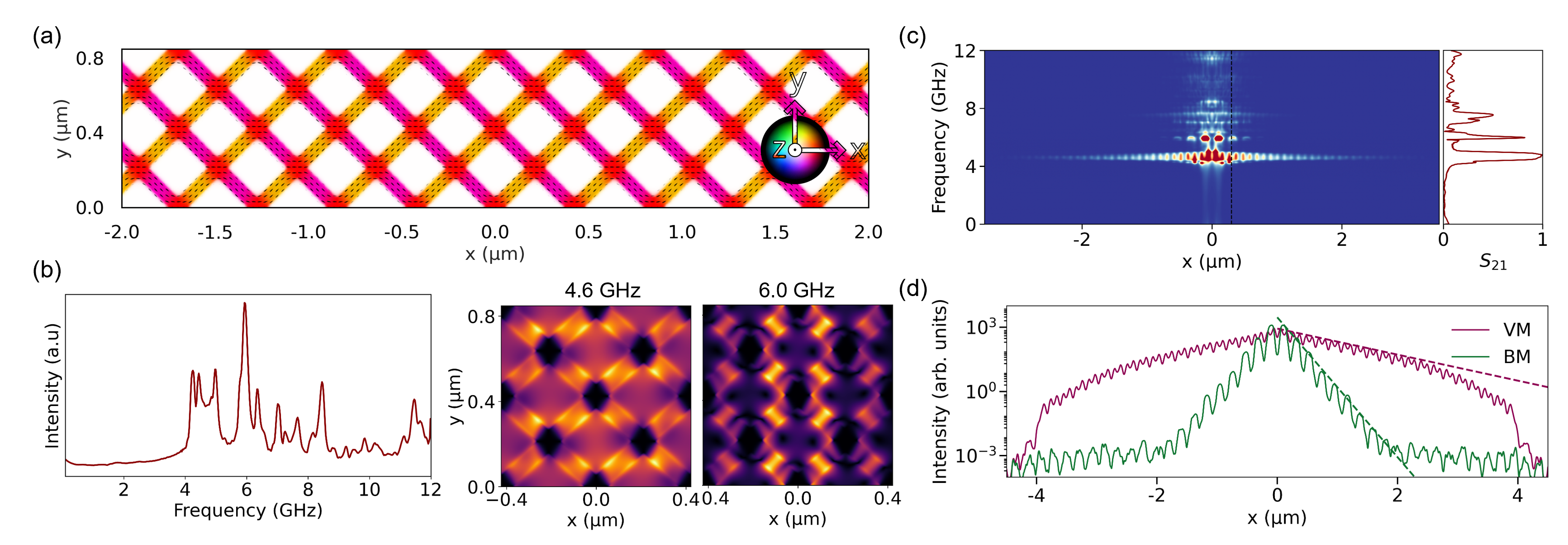}
    \caption{\textbf (a) Relaxed static magnetization configuration in c-ASI lattice. White regions: PMA matrix magnetized out-of-the-plane. (b) Normalized SW spectra and mode profiles of vertex mode (VM, 4.6 GHz) and BM, 6.0 GHz). (c) Left: 2D color map of the averaged (along the $y$-axis) intensity of SWs in their propagation along $\pm x$ direction, Right: Normalized transmission spectra ($S_{21}$) measured 0.3~µm from antenna. (d) Intensity of VM and BM along $\pm x$-directions on log scale. Dotted lines: exponential fits $\propto \exp(-2x/\xi)$.}
    \label{figS8}
\end{figure*}

Simulations were also performed on interconnected ASI-PMA under identical conditions (Fig.~\ref{figS8}(a)). A distinct vertex mode (VM) emerges at 4.6 GHz with amplitude localized at junctions (Fig.~\ref{figS8}(b)), alongside BM at 6.0 GHz. Spatial decay profiles (Fig.~\ref{figS8}(c,d)) reveal that VM propagates over ~1.4 µm compared to ~0.3 µm for BM. This enhanced VM propagation partially addresses a key limitation of c-ASI: domain-wall pinning and interface scattering that typically suppress SW transmission~\cite{sahoo2021observation, may2021magnetic}. However, compared to disconnected ASI, the connected ASI-PMA configuration offers less mode diversity and tunability, making the disconnected geometry more suitable for reconfigurable magnonic applications.

\bibliographystyle{apsrev4-2}
\bibliography{bibliography_supp}